\documentclass[12pt]{iopart}
\usepackage{iopams}
\usepackage{graphicx}
\usepackage{amsfonts}

\begin{document}
\title{Josephson current through interacting double quantum dots with spin-orbit coupling}

\author{Stephanie Droste$^{1,2}$, Sabine Andergassen$^{1,3}$, Janine Splettstoesser$^1$} 
\address{$^1$Institut f\"ur Theorie der Statistischen Physik, RWTH Aachen University, 52056 Aachen, Germany and JARA-Fundamentals of Future Information Technology}
 \address{$^2$School of Chemical and Physical Sciences and MacDiarmid Institute for Advanced Materials and Nanotechnology, Victoria University of Wellington, PO Box 600, Wellington 6140, New Zealand}
\address{$^3$Faculty of Physics, University of Vienna, Boltzmanngasse 5, 1090 Vienna, Austria}
\ead{stephanie.droste@vuw.ac.nz}
\begin{abstract}
We study the effect of  Rashba spin-orbit interaction on the Josephson current through a double quantum dot in presence of Coulomb repulsion. In particular, we describe the characteristic effects on the magnetic-field induced singlet-triplet transition in the molecular regime. Exploring the whole parameter space, we analyze the effects of
the device asymmetry, the orientation of the applied magnetic field with respect to the spin-orbit interaction, and 
finite temperatures. 
We find that at finite temperatures the orthogonal component of the spin-orbit interaction exhibits a similar effect as the Coulomb interaction inducing the occurrence of a $\pi$-phase at particle-hole symmetry. This provides a new route to the experimental observability of the $\pi$-phase in multi-level quantum dots.
\end{abstract}


\submitto{\JPCM}
\maketitle
\section{Introduction}
The Josephson effect is at the basis of Cooper pair transport through a weak link between two superconducting contacts. A current can flow due to the phase difference between the superconducting condensates; depending on the sign of the resulting current, one speaks of the $0$- or the $\pi$-phase. The study of the Josephson effect is particularly insightful, if the junction itself has an internal structure, where the Josephson current is carried due to the formation of Andreev bound states. In quantum-dot structures a large tunability of the quantum-dot junction by electric gating or externally applied magnetic fields is provided. Recently  
quantum-dot superconductor hybrid structures (see Ref.~\cite{deFranceschi10, Rodero11} for a review)
have been realized and the Josephson current as well as Andreev bound states have been studied in a systematic and controlled way~\cite{Buitelaar02,Eichler07,Jespersen07,Grove07,Eichler09,Cleuziou06,Maurand12,vanDam06,Buitelaar03,Jarillo06,Jorgensen06,Jorgensen07,Jorgensen09,Pillet10,Hofstetter09,Herrmann10}. 
Whether the Josephson current through the system contributes in the $0$- or the $\pi$-phase  or whether it is completely suppressed, depends on the internal parameters of the quantum dot. In particular the Coulomb interaction has been shown to strongly affect the shape of the Josephson current~\cite{vanDam06,Governale08,Futterer09,Karrasch11,Meng09,Vecino03,Luitz10,Zazunov09,Choi00}, e.g. through the Kondo effect~\cite{Buitelaar02,Eichler07,Jespersen07,Grove07,Eichler09,Cleuziou06,Maurand12,Lopez07,Lee10,Zitko10,Zazunov10,Choi04,Oguri04,
Siano04,Siano04E,Tanaka07,Bauer07,Hecht08,Karrasch08}. Theoretical studies dealt with the single-level Anderson impurity 
coupled to BCS leads as a minimal 
model for the analysis of phase boundaries and the related transition in the Josephson 
current~\cite{Meng09,Vecino03,Luitz10,Choi04,Oguri04,Siano04,Siano04E,Tanaka07,Bauer07,Hecht08,Karrasch08,Glazman89,Rozhkov99,Novotny05,Luitz12}.

\begin{figure}[b]
\center{\includegraphics[width=2.7in,clip=true]{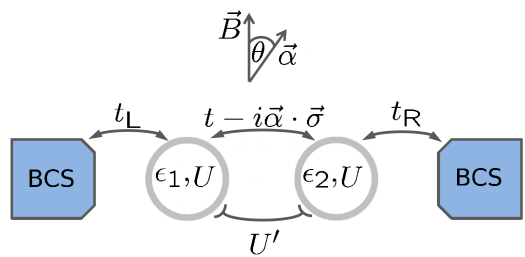}}
\caption{(Color online) Sketch of two serially coupled  quantum dots with different energy levels $\epsilon_{1/2}$ and onsite $U$, as well as  interdot Coulomb interaction $U'$. The dots are coupled to each other by an interdot coupling with amplitude $t$, as well as by a spin-orbit interaction $\vec{\alpha}$. The external magnetic field $\vec{B}$ spans an angle $\theta$ with respect to the SO direction. The double quantum dot is connected to superconducting leads, with gap $\Delta \,e^{i\phi_{L/R}}$, by a tunneling coupling $t_\mathrm{L/R}$.
\label{fig:model}
}
\end{figure}

Various experimental realizations of Josephson multi-level quantum dots include carbon nanotubes, in addition to other materials such as InAs, both exhibiting strong spin-orbit (SO) effects~\cite{Deacon10}.
While for the latter the relevance of the SO interaction is known~\cite{Winkler03}, for carbon nanotubes it has only lately been attributed
to the curvature of the tube~\cite{Ando00,Kuemmeth08,Jespersen11}.
SO coupling has an important impact on spin lifetimes in quantum dots~\cite{Khaetskii00,Fujisawa02,Elzerman04}. Several SO interaction effects on transport through quantum dots with normal leads have furthermore been studied, as e.g. the impact on weak localization and universal conductance fluctuations~\cite{Aleiner01}, the suppression of the Kondo ridges in combination with a Zeeman field~\cite{Grap11}, and the occurrence of spin-polarized currents~\cite{DelValle11}.
Observations of this type trigger the need to study the SO interaction on the Josephson current through interacting quantum dots. This is of particular relevance regarding the singlet character of the Cooper pairs carrying the Josephson current and the resulting sensitivity to interactions breaking the spin symmetry.
Lately, different aspects of the SO interaction on the Josephson current and Andreev bound states have been considered in the regime of vanishing Coulomb interaction with an emphasis on the quasiparticle contributions to the current~\cite{Zazunov09,Dolcini08,Yang08,Beri08,DellAnna07,Chtchelkatchev03,Krive04,Bezuglyi02,Cheng12}.

In this manuscript, we focus on the effect of spin-orbit interaction on the  Josephson current through quantum dots, allowing for a strong Coulomb interaction, where a magnetic field with arbitrary orientation with respect to the spin-orbit field can be applied. We thoroughly take into account the variety of parameters, governing a realistic quantum dot setup such as asymmetries in the coupling due to gating and contrast their properties with the characteristics of the SO interaction. We perform this study for zero as well as for finite temperatures. 

In order to assess the role of SO interactions in multi-level structures we here focus on a double quantum dot as a paradigm system, where the localized dots represent the different orbits. It has been shown in Refs.~\cite{Karrasch11,Meng09}, that the main features in the Josephson current through a single or a double dot setup can already be detected in the molecular regime of infinitely large  superconducting gaps $\Delta$. We therefore concentrate on the regime of $\Delta\rightarrow\infty$, which describes well the subgap features. In this regime single-particle transport is suppressed and the full spectrum can be accessed.  Our detailed study can serve for the characterization of quantum-dot Josephson junctions and provides a reference for studies including quasiparticle transport as well~\cite{Zazunov09,Choi00,Zazunov10,Lim11}.

The paper is organized as follows. In Sec.~\ref{sec:model}, we introduce the double-dot 
model and present the full effective Hamiltonian. In Sec.~\ref{sec:res} we discuss the results for the Josephson current,
starting from the symmetric double quantum dot, subsequently addressing the effects of the Coulomb interaction, SO interaction, finite asymmetries and finite temperatures.
We finally extend this analysis to the more general case of a parallel quantum-dot configuration, where we focus on a special transport regime
which cannot be accessed by the serial geometry.

\section{Serial quantum dot model}
\label{sec:model}
\subsection{Microscopic Hamiltonian}
We consider two serially coupled quantum dots with a single electronic orbital in each dot contributing to transport. All hopping amplitudes in the model are assumed to be real, except for the imaginary hopping amplitude $i\alpha$ taking into account the Rashba SO interaction. 
The model we use is 
sketched in  Fig.~\ref{fig:model}. The Hamiltonian of the isolated double dot is given by
\begin{equation}
H_{\rm dd}=H_0+H_{\rm SO}+H_{\rm int} \ ,\label{eq_Hdot}
\end{equation}
containing the free part $H_0$, a term due to the SO coupling $H_{\rm SO}$, and the Coulomb interaction contributions, $H_{\rm int}$. The free part is
\begin{equation}
H_0=\sum_{\sigma} \left[\sum_{j=1,2} \epsilon^{}_{j,\sigma} d_{j,\sigma}^\dagger
  d_{j,\sigma}^{\phantom{\dagger}}-\frac{t}{2}\left(
d_{2,\sigma}^\dagger d_{1,\sigma}^{\phantom{\dagger}} + d_{1,\sigma}^\dagger d_{2,\sigma}^{\phantom{\dagger}}\right)\right] \; ,
\end{equation}
with $d_{j,\sigma}^\dagger$ being the creation operator of an electron
on the dot site $j=1,2$ with spin $\sigma=\uparrow,\downarrow$. The on-site energies of the two quantum dots are 
\begin{equation}
\epsilon_{1/2,\sigma}=-\epsilon +\sigma\frac{B}{2} \pm \frac{\delta}{2}\ ,
\end{equation} 
which can be tuned by an external gate voltage and a magnetic field $\boldsymbol{B}$ with $|\boldsymbol{B}|=B$, lifting the spin degeneracy. (We here choose $\mu_\mathrm{B}=e=\hbar=k_\mathrm{B}=1$.)
 The resulting Zeeman splitting is given by $B$ with $\sigma=\pm$ for spin up and down respectively; we neglect the effect of the magnetic field on the orbital motion. The Zeeman field $\boldsymbol{B}=B\boldsymbol{e}_z$ 
 sets the spin quantization axis, which we here choose to point along the $z$-direction. The difference between the orbital on-site energies of the two dots is parametrized 
by the level detuning $\delta=\epsilon_{1,\sigma}-\epsilon_{2,\sigma}$. 
Hybridization of the single-dot states occurs due to the spin-independent  interdot hopping with amplitude $t$.

The Rashba SO interaction we are interested in is taken into account by 
an imaginary hopping amplitude of spin-dependent sign as the 
lattice realization resulting from spatial 
confinement in semiconductor structures~\cite{Winkler03,Mireles}. Importantly, it also applies for the curvature-induced SO interaction which occurs in carbon nanotubes~\cite{Ando00,Kuemmeth08,Jespersen11}.
The Rashba term of the Hamiltonian with $\alpha>0$  reads
\begin{equation}\label{eq:SOI}
H_{\rm SO}  =  i \alpha \sum_{\sigma,\sigma'} \left[  
d_{2,\sigma}^\dag (\sigma_z)_{\sigma,\sigma'} d_{1,\sigma'} {\rm cos} \, \theta + 
d_{2,\sigma}^\dag (\sigma_x)_{\sigma,\sigma'} d_{1,\sigma'} {\rm sin} \, \theta \right] +\mbox{H.c.} \; ,
\end{equation}
where $\theta$ is the angle between the effective SO field $\boldsymbol{\alpha}$ with $|\boldsymbol{\alpha}|=\alpha$ and the Zeeman field $\boldsymbol{B}$. The Pauli matrices $\sigma_{x}$ and $\sigma_{z}$ are related to the electron spin.  
The SO interaction can have a component parallel to $\boldsymbol{B}$ in
$z$-direction and a perpendicular one, which - without loss of generality - 
we choose to be parallel to the $x$-axis. 
For $\theta = \pm \pi/2$ the SO interaction and the $B$-field are orthogonal, while they are parallel for 
$\theta=0$. It can be shown that in the latter case the conventional hopping and the SO interaction can be 
combined to an effective hopping $\tilde{t}=\sqrt{t^2+4\alpha^2}$. 

The local Coulomb repulsion between electrons occupying the double-dot is modeled as an on-site interaction $U$ as well as a nearest-neighbor interaction $U'$
\begin{equation}
H_{\rm int}=U\sum_{j=1,2} \left(n_{j,\uparrow} - \frac{1}{2} \right) \left( n_{j,\downarrow} - \frac{1}{2} \right) +U' \left( n_{1} -1 \right) \left(n_{2}-1\right) 
\label{eq:int}
\end{equation}
 where the number of electrons on each dot is $n_j= \sum_\sigma n_{j,\sigma}$, with $n_{j,\sigma}= d_{j,\sigma}^\dagger  d_{j,\sigma}^{\phantom{\dagger}}$. 
Our model Hamiltonian is chosen such that $\epsilon=0$ corresponds to half filling. 

We consider the double-dot structure described above to be coupled to two superconducting leads. The left and right lead are modeled by the BCS Hamiltonian 
\begin{equation}
H_\mathrm{lead}^{s=L,R}=
\sum_{k,\sigma}\epsilon_{k}c^\dagger_{sk\sigma}c_{sk\sigma}^{\phantom{\dagger}} 
- \Delta \sum_k\left(e^{i\phi_s}c^\dagger_{sk\uparrow}c^\dagger_{s-k\downarrow} + \mathrm{H.c.}\right)~,\label{eq_Hleads}
\end{equation}
with $\Delta$ and $\phi_{L,R}=\pm\phi/2$ being the BCS gap and phase, respectively. 
The coupling to the leads is modeled by the tunnel Hamiltonian
\begin{equation}
H_\mathrm{T} = \sum_{k,\sigma}\left[t_\mathrm{L} c^\dagger_{\mathrm{L}k\sigma}d^{\phantom{\dagger}}_{1,\sigma}+t_\mathrm{R} c^\dagger_{\mathrm{R}k\sigma}d^{\phantom{\dagger}}_{2,\sigma} +\mathrm{H.c.}\right] \ .
\end{equation}
We furthermore assume the density of 
states $\rho_s$ in the leads in the normal state to be constant (this is particularly reasonable since we later assume large superconducting gaps). In this  wide-band limit, the coupling strength to the leads is defined as
\begin{equation}
\Gamma_s=\pi t_s^2\rho_s ~.
\end{equation}
The total coupling is given by $\Gamma=\Gamma_\mathrm{L}+\Gamma_\mathrm{R}$ and we define the asymmetry in the coupling strengths to the two leads by $\beta=(\Gamma_\mathrm{L}-\Gamma_\mathrm{R})/\Gamma$.

\subsection{Effective Hamiltonian for $\Delta \to \infty$}\label{sec:effective}

We treat the problem by introducing an effective Hamiltonian in the limit where the superconducting gap is the largest energy scale. This allows to analytically determine the many-body eigenstates of the system as a starting point to exactly compute the Josephson current. 

We therefore set up the equations of motion for the double-dot retarded Green's functions $\hat{G}^\mathrm{ret}_{ji}(t)=-i\theta(t)\langle \left\{\Psi_i(t),\Psi_j^\dagger(0)\right\}\rangle$, using the Hamiltonian introduced in the section above. 
  Here, the so-called Nambu spinor of dot $i$ is given by $\Psi_i=( d^{}_{i\uparrow },d^{\dagger}_{i\downarrow }) $.
The superconductor couples electrons and holes, while the particle number is not fixed, due to fluctuations in the superconducting condensate. Therefore it is useful to introduce the Nambu basis. This means that we count electrons of spin $\sigma$ and missing electrons of opposite spin $\bar{\sigma}$ as Nambu particles. Importantly, the Nambu particle number is fixed in the presence of particle transfer due to Andreev reflection only. Due to the coupling to the superconducting leads also off-diagonal elements in this Green's function appear.

In the limit $\Delta \to \infty$, the same set of equations of motion for $\hat{G}^\mathrm{ret}_{ji}$ can be reproduced by an effective dot Hamiltonian, see the Appendix for details. For the double dot described in the previous section, this effective Hamiltonian is given by
\begin{eqnarray}
\label{eq:hatom}
H_\mathrm{eff} = H_0+H_\mathrm{SO}+H_\mathrm{int}+ \left(\Gamma_\mathrm{L} e^{-i\phi/2} d_{1\uparrow}d_{1\downarrow} + \Gamma_\mathrm{R}e^{i\phi/2} d_{2\uparrow}d_{2\downarrow} + \mathrm{H.c.}\right)~.
\end{eqnarray}

Importantly, the proximity of the superconducting leads introduces a term which couples electrons and holes of opposite spin on each dot by Andreev reflection, but no term coupling particles on different dots is induced. 
We now individuate the relevant Nambu particle subspaces of the spectrum and their role for the Josephson current through the device.
By introducing Nambu spinors, we can decompose the 16-dimensional Hilbert space of the proximized double quantum dot that 
underlies $H_\mathrm{eff}$ into uncoupled sectors with different Nambu-particle numbers, $\{0,\ldots,4\}$. We note that the Coulomb interaction enters in a non-trivial way only the sector with Nambu-particle number 2.
For the analysis of the ground state and the resulting Josephson current, it is helpful to relate the eigenstates of the effective Hamiltonian with a fixed Nambu-particle number with the total spin $s$ and the $z$-component $s_z$. The Nambu-particle number is given by $2s_z+2$.

The characterization of the subspaces by the $z$-component of the  spin holds only in the absence of SO interaction or as long as the SO interaction is parallel to the magnetic field $\boldsymbol\alpha\|\boldsymbol{B}$. For any finite orthogonal component $\alpha_{\bot}$ with respect to the magnetic field, states with a difference $\left|\Delta s_z\right|=1$ of the spin $z$-component couple due to SO interaction-induced spin flips. The result is that only two independent subspaces are formed, characterized by an even, respectively an odd, Nambu-particle number. This coupling of different Nambu subspaces, will be shown to influence importantly the Josephson current.
   
We are interested in the Josephson current through the device, which we can access through the full many-particle spectrum. Having calculated the set of eigenvalues $E_i$, the Josephson current at finite temperature can be obtained from the phase derivative of the 
free energy $F$~\cite{Bloch70,Beenakker92} by
 \begin{equation}\label{eq:Jos}
J = 2\partial_\phi F~,~~~F=-T\ln\sum_i e^{-E_i/T}
\end{equation}
with $F=-T\ln\sum_i e^{-E_i/T}$.
This equation shows that in principle all eigenstates of the double quantum dot contribute to the total Josephson current with a contribution weighted by their energy.
At zero temperature, $T=0$, only the many-particle groundstate $E_0$ contributes to the Josephson current. Therefore Eq.~\eref{eq:Jos} reduces to
\begin{equation}\label{eq:JosEg}
J = 2\partial_\phi E_0~.
\end{equation}
It depends on the parameters of the system, which subspace the many-particle ground state stems from. This can be tuned by a magnetic field $B$ and the gate voltage entering the level positions $\epsilon_j$. 

In the following section, we delineate the effect of the Coulomb interaction on the Josephson current   in the double quantum dot introduced above, utilizing this as a basis to study in detail the effect of  SO interaction and asymmetries. 

\section{Results}
\label{sec:res}

Based on the energy spectrum calculated from the effective Hamiltonian, Eq.~(\ref{eq:hatom}), we present results for  the Josephson current in the following sections. 
We choose the coupling between the dots and between dots and leads to be of the same order $t\simeq\Gamma$, except when otherwise indicated. In this regime, the eigenstates of the isolated double quantum dot are given by the molecular bonding and antibonding states, namely by the symmetric and antisymmetric superposition of the local quantum-dot states. Furthermore,  we consider $U=U'$ for simplicity, which is a reasonable choice for the double dot being in the molecular regime.

\subsection{Effect of Coulomb interaction}\label{sec:coulomb}

It has been shown before that for a symmetric double-dot setup without SO interaction and for sufficiently large Coulomb interaction, a singlet-triplet transition occurs as a function of an externally applied magnetic field resulting in a discontinuity in the Josephson current at a critical magnetic field~\cite{Karrasch11}. We compare our results to this basic symmetric case and discuss in detail the impact of SO interaction, coupling and gate-voltage asymmetries and finite temperatures.

\begin{figure}[t]
\center
\includegraphics[width=4in,clip=true]{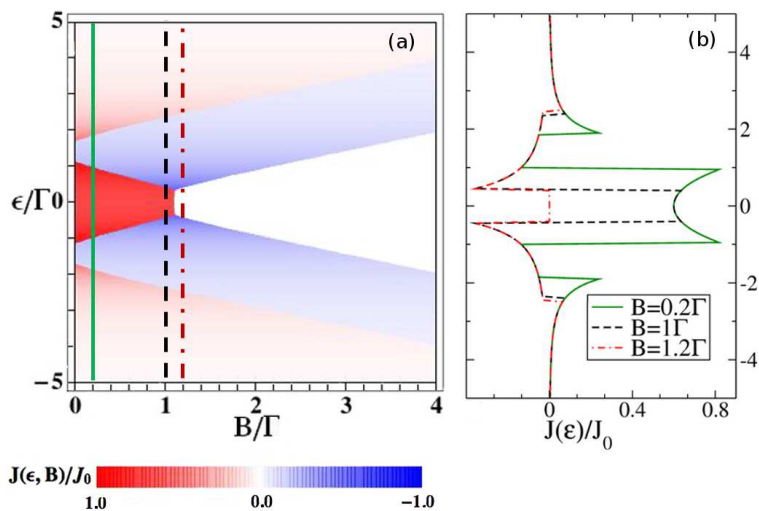}
\caption{(Color online) (a) Josephson current for the symmetric ($\beta=\delta=0$), interacting double dot with  $U=\Gamma$ as a function of the magnetic field $B$  and the mean level position $\epsilon$ in units of $\Gamma$. (b) Cuts of the density plot in (a) at fixed values of $B$, as a function of $\epsilon$. The other parameters are $t=\Gamma$, $k_\mathrm{B} T=0$, 
and $\phi=\pi/2$.
\label{fig:J_generic}}
\end{figure}

We  start our discussion by considering the effect of the Coulomb interaction on a fully symmetric dot, $\delta=\beta=0$, in the absence of SO interaction ($\alpha=0$) and  at zero temperature, $k_\mathrm{B}T=0$. The Hamiltonian is symmetric with respect to $B \to -B$ and  we therefore restrict our analysis to $B>0$. 
In Fig.~\ref{fig:J_generic}, we show the result for the Josephson current through the double quantum dot, $J(\epsilon,B)$, as a function of the mean level position, which can be tuned by applying a gate voltage, and as a function of an external magnetic field, at $U=\Gamma$. This generic scenario for the regime of large Coulomb interaction $U>U_c$ (with a critical interaction $U_c$ to be addressed later) has been discussed in Ref.~\cite{Karrasch11}. 

We show the Josephson current  in units of a reference Josephson current at $\epsilon=0$, which flows through a fully symmetric and non-interacting quantum dot, $U=\delta=\beta=\alpha=0$, at zero magnetic field and temperature, $B=k_B T=0$; for the plots shown in this paper it takes the value $J_0(t=\Gamma,\Gamma,\phi=\pi/2)=\Gamma/(2\sqrt{2+\sqrt{2}})$, except when otherwise indicated.

The Josephson current $J(\epsilon,B)$ as displayed in Fig.~\ref{fig:J_generic} features three different phases, namely the $0$-phase with $J>0$ (for $0<\phi<\pi$), the $\pi$-phase characterized by $J<0$ (for $0<\phi<\pi$), and the triplet phase with $J=0$.
The respective ground state for the $0$-phase is a non-degenerate singlet with $\left\{s=0,s_{z}=0\right\}$ (Nambu particle number $2$). For the $\pi$-phase the ground state is associated with a free spin~\cite{vanDam06,Meng09,Glazman89,Spivak91,Baselmans99,Rozhkov01,Ryazanov01}; at $B=0$, this is given by the degenerate doublet with
$\left\{s=1/2,s_z=\pm1/2\right\}$ (namely Nambu-particle number $1$ or $3$). The triplet state (Nambu-particle number $0$ or $4$) has $\left\{s=1,s_z=\pm1\right\}$, respectively.
We furthermore note that $J(\epsilon,B)$ is symmetric with respect to $\epsilon \to -\epsilon$. The symmetry in $\epsilon$ is preserved as long as either $\delta=0$ or $\beta=0$ holds.
 
In the vicinity of $B\simeq0$, depending on the gate voltage $\epsilon$, a positive Josephson current can flow through the double quantum dot in the $0$-phase, 
associated with a non-degenerate spin singlet, see red regions in Fig.~\ref{fig:J_generic}. 
However, due to the finite Coulomb interaction suppressing double occupation, a gate-voltage regime develops in which 
one of the molecular bonding and antibonding states is occupied with a single spin, leading to a \textit{reversal of the Josephson current}. This latter is associated to a $\pi$-phase, where the double-dot state is in an almost twofold degenerate state with a free spin (exactly degenerate at $B=0$). These two $0$-$\pi$-$0$ transitions as a function of $\epsilon$, occurring due to the single-level behavior, are observed as long as $B<B_c$, where the critical magnetic field $B_c$ is defined by a degeneracy of a bonding and an antibonding state with opposite spin~\cite{Meng09,Vecino03,Luitz10,Choi04,Oguri04,Siano04,Siano04E,Tanaka07,Bauer07,Hecht08,Karrasch08,Glazman89,Rozhkov99,Novotny05,Luitz12}; this degeneracy corresponds to a degenerate singlet and triplet ground state.

A finite Zeeman field $B$ splits the spin degeneracy in the
$\pi$-phase, leading at the same time to a broadening of the regions
with negative Josephson current. The width of these regions depends on
the strength of the coupling to the superconducting leads $\Gamma$ and
their phase difference $\phi$ and is of the order of $U+B_c$ for $B>B_c$.
The singlet remains unaffected by the magnetic field.  
For $B\lesssim B_c \sim t$, the $\pi$-phase behavior is much more prominent than the singlet behavior close to zero gate voltage. Instead of a twofold $0$-$\pi$-$0$ transition one would now rather speak of a single $\pi$-$0$-$\pi$ transition as a function of the gate voltage. The resemblance of these features of opposite sign, at $B\simeq 0$ and $B\lesssim B_{c}$, regarding the gate-voltage characteristics of the current $J(\epsilon)$ 
as well as the current-phase relation $J(\phi)$, can be perceived from panel (b) in Fig.~\ref{fig:J_generic} and is discussed in detail in Ref.~\cite{Karrasch11}. 
Moreover, the line-shape of $J(\epsilon)$ for $B\simeq B_c$ is characterized by a discontinuity 
as a function of the magnetic field across the singlet-triplet transition at $B_c$, as illustrated in Fig.~\ref{fig:J_generic}. The suppression of $J$ in the triplet spin configuration, where the occupation of the two levels with equal spin is favored, is due to the inhibited Cooper-pair tunneling with respect to the singlet one~\footnote{This complete suppression of the Josephson current turns into a strongly reduced, but finite Josephson current in presence of quasi-particle transport occurring for a finite superconducting gap. However, these corrections turn out to be small as long as $\Delta\gg\Gamma$.} As a function of $\epsilon$, the width of the discontinuous singlet-triplet transition is of the order of the Coulomb interaction $U$ for the symmetric case (as for the $\pi$-phase at $B=0$). 
 The occurrence of a singlet-triplet transition is obviously unique to a multi-orbital system.
 
\begin{figure}[t]
\center
\includegraphics[width=3.5in]{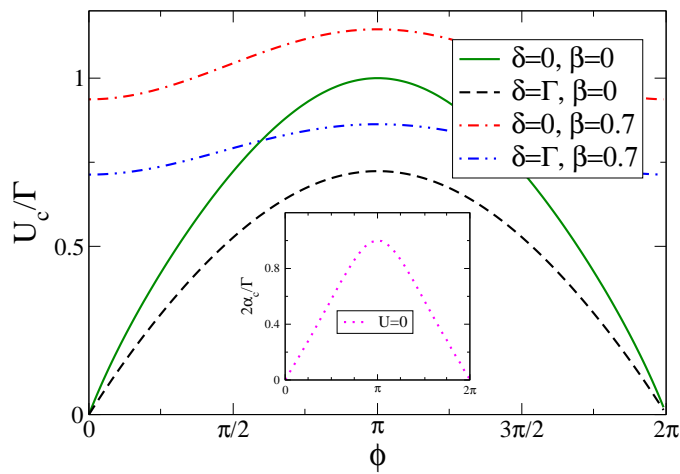}
\caption{(Color online) Critical Coulomb interaction $U_c$  as a function of the superconducting phase difference $\phi$. The other parameters are $t=\Gamma$, and $k_B T=0$. Inset: Critical SO coupling strength $2\alpha_c$ as a function of the superconducting phase difference $\phi$ at $U=0$. The other parameters are $\theta=\pi/2$, $t=\Gamma$, and $k_B T=0$. 
\label{fig:Uc}}
\end{figure}

The discontinuity in $J$ as a function of $B$ at a critical magnetic field $B_c$ occurs only for sufficiently large Coulomb interaction ($U>U_{c}$), where the critical interaction $U_c$ depends on the superconducting phase difference, see the full green line in Fig.~\ref{fig:Uc}. We observe that for  small values of $\phi$, the singlet-triplet transition is observable already for rather weakly interacting quantum dots. For $U=0$ and $\phi\rightarrow0$ all three phases, $0$, $\pi$ and triplet, touch at $\epsilon=0$ and $B=B_c$. In this case the discontinuous singlet-triplet transition as a function of $B$ occurs hence only for $\epsilon=0$.

\begin{figure}[t]
\center
\includegraphics[width=3.5in]{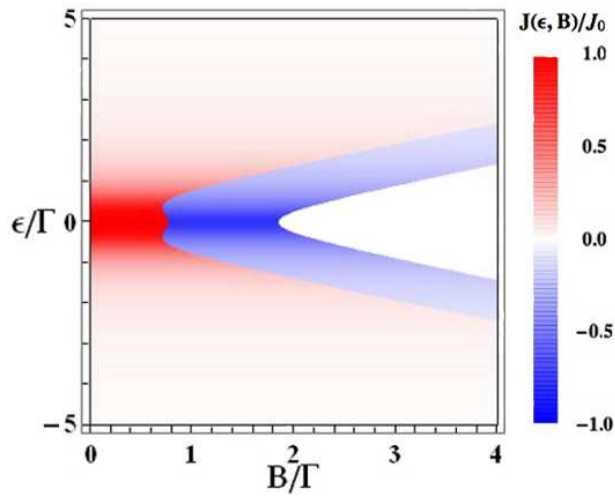}
\caption{(Color online)  Josephson current as a function 
of the magnetic field and the level position in units of $\Gamma$ for $U=0$ ($U<U_\mathrm{c}$). The other parameters are $\delta=\beta=0$, $t=\Gamma$, $k_B T=0$, and  $\phi=\pi/2$.
\label{fig:Jos_sym}}
\end{figure}

The main effect of the Coulomb interaction being smaller than the critical one, $U<U_c$, can be observed in Fig.~\ref{fig:Jos_sym}, exemplified for the non-interacting, symmetric double dot. Note that the result remains qualitatively the same for any $U<U_c$. 
The direct singlet-triplet transition, which for large $U$ appears at $B=B_c$, see Fig.~\ref{fig:J_generic}, is inhibited by an interjacent $\pi$-phase, extending between the $0$- and the triplet phase.  
The reason for this is that for weak Coulomb interaction and a magnetic field of the order of the single level spacing, a non-degenerate ground state with a magnetic moment associated to a spin $1/2$ can exist at $\epsilon\simeq0$. The maximum strength of the Coulomb interaction allowing for the existence of such a spin-$1/2$ ground state at $\epsilon\simeq0$ is given by the critical interaction $U_c$.

We now focus on  the $B$-field interval giving rise to the $\pi$-phase at $\epsilon=0$, which prevents the direct transition between the singlet and the triplet phase. This interval is given by
\begin{eqnarray}
\delta B &=& 2\sqrt{t^{2}+\Gamma^{2}+\sqrt{2t^{2}\Gamma^{2}\left(1-\cos\phi\right)}}- \nonumber
\\&& \sqrt{2}\sqrt{t^{2}+\Gamma^{2}+\sqrt{t^{4}+\Gamma^{4}+2t^{2}\Gamma^{2}\cos\phi}}\; .
\label{eq:B_sym_distance}
\end{eqnarray}
We will discuss the effect of SO interaction and of asymmetries in the following subsection. From Eq.~(\ref{eq:B_sym_distance}) we see that for $\phi\rightarrow 0$ the $B$-field interval, in which the $\pi$-phase occurs at $\epsilon=0$ vanishes. This means that at $\phi\rightarrow 0$ a direct singlet-triplet transition as a function of $B$ occurs at $\epsilon=0$ already for vanishing Coulomb interaction. This finding is confirmed by the solid green line in the plot of Fig.~\ref{fig:Uc}, where the critical Coulomb interaction $U_c$ is zero at $\phi=0$. 

A further effect of the Coulomb interaction is that it tends to decrease the absolute value of the Josephson current. This can be seen from a comparison of Figs.~\ref{fig:J_generic} and \ref{fig:Jos_sym}. It becomes even clearer in the quantitative comparison given by the solid green lines for the Josephson current as a function of the magnetic field at $\epsilon=0$ in the plots shown for $U=0$ and $U=\Gamma$, in Fig.~\ref{fig:Joscut_SOI}. 
\subsection{Spin-orbit interaction effects}\label{sec:SOI}

\begin{figure}[h]
\center
\includegraphics[width=3.5in]{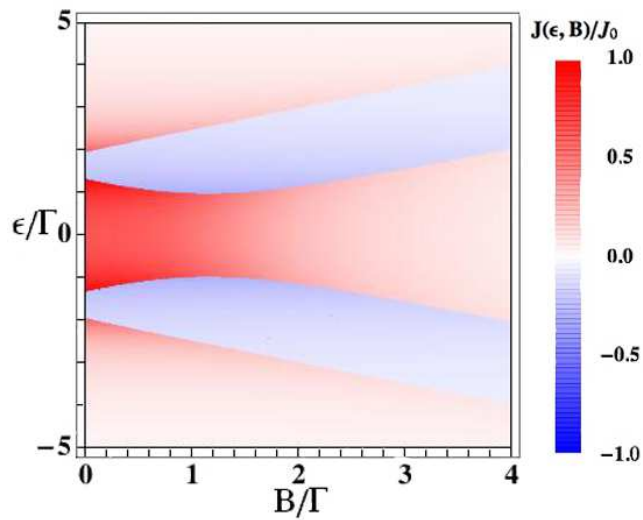}
\caption{(Color online) Josephson current for the symmetric ($\delta=\beta=0$), interacting case, $U=\Gamma$, as a 
function of the magnetic field and the level position in units of $\Gamma$. The angle between the magnetic field and the SO interaction field is $\theta=\pi/2$. 
Furthermore $\alpha=0.5 \Gamma$, $t=\Gamma$, $k_B T=0$, and $\phi=\pi/2$.
\label{fig:J_SOI}}
\end{figure}

We now come to the main purpose of this paper, namely the study of SO effects. The SO interaction breaks spin-rotational symmetry by designating a  certain (spin) direction. Spin is hence no longer a good quantum number, however at zero magnetic field 
a Kramers doublet remains as time-reversal symmetry is conserved. 
In the presence of an additional, externally applied finite Zeeman field, the direction of this external field with respect to the effective SO field is important. In the Hamiltonian responsible for the SO interaction in the double-dot setup, Eq.~(\ref{eq:SOI}), the angle $\theta$ between the SO interaction and the $B$ field defines their respective orientation. A parallel configuration is similar to the situation discussed before. In this case, the amplitude of the SO interaction affects only the interdot tunneling, which takes a SO interaction-dependent effective value, $\tilde{t}=\sqrt{t^2+4\alpha^2}$, and therefore the results of Sec.~\ref{sec:coulomb} hold.

\begin{figure}[h]
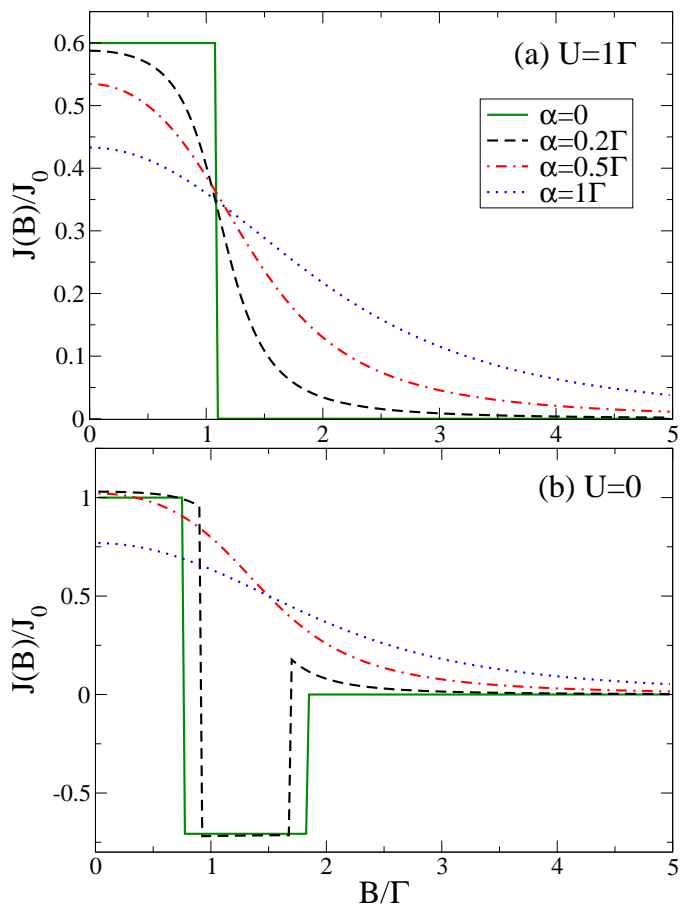

\centering
\includegraphics[width=3.5in]{fig6a}
\centering
\includegraphics[width=3.5in]{fig6b}
\caption{ (Color online) Josephson current for $\epsilon=0$ and different values of the SO coupling 
strength $\alpha$ as a function of the magnetic field in units of 
$\Gamma$. We show the symmetric case, $\delta=\beta=0$, for a Coulomb interaction of (a) $U=\Gamma$ and (b) 
$U=0$. The angle between the magnetic field and the SO interaction field is $\theta=\pi/2$. Furthermore 
$t=\Gamma$, $k_B T=0$, and $\phi=\pi/2$.
\label{fig:Joscut_SOI}}
\end{figure}

 In the following we focus on the
effect of a finite orthogonal component $\alpha_{\perp}$ and therefore concentrate on the case $\theta = \pi/2$ with $\alpha=\alpha_\perp$.
A finite orthogonal component of the SO interaction with respect to the Zeeman field 
introduces an anticrossing at finite $B$ of the levels with a finite spin in either one of the dots or in both and consequently it leads to an avoided singlet-triplet transition, influencing the previously discussed discontinuity in the Josephson current as a function of $B$. 
For simplicity we consider the symmetric case with $\beta = 0$ and $\delta=0$ and discuss the asymmetric case in the subsequent section. The effect of the SO interaction on the Josephson current is shown in Fig.~\ref{fig:J_SOI}, where we plot the Josephson current $J(\epsilon,B)$ for $\alpha_\perp= 0.5\Gamma$ and $U>U_{c}$.
The singlet-triplet transition as a function of $B$ leading to a discontinuity in the Josephson current for $\alpha_{\perp}=0$, as observed in Fig.~\ref{fig:J_generic}, turns into a smooth crossover for a finite $\alpha_\perp$ between the $0$-phase and the triplet phase, where the Josephson current is strongly suppressed only in the limit of large magnetic fields. To show this effect quantitatively, we plot in Fig.~\ref{fig:Joscut_SOI}~(a) the Josephson current for $\epsilon=0$ as a function of $B$ for various values of $\alpha_{\perp}$, clearly demonstrating the smooth crossover introduced by the perpendicular component of the SO interaction field. 
For comparison, in Fig.~\ref{fig:Joscut_SOI}~(b) we show the corresponding non-interacting result, $U=0$. 
For small SO interaction a direct singlet-triplet crossover is inhibited by an interjacent $\pi$-phase, see also Fig.~\ref{fig:Jos_sym}. However, for $\alpha_{\perp}$ larger than some critical value $\alpha_c$, the $\pi$-phase is again not accessible at $\epsilon\simeq0$ and the full (smoothened) crossover line between singlet and triplet phase is recovered. The Josephson current is hence positive for all $B$ at $\epsilon=0$. In this sense the SO interaction has a similar effect as the Coulomb interaction: it opens a window of positive Josephson current between the two $\pi$-phases, which appear due to transport when one of the different molecular states is occupied with a half-integer spin. 
This can also be understood by analyzing the relevant eigenstates of the system.
The gap opened by the SO interaction in the single-particle energy spectrum of the isolated double dot
leading to the anti-crossing is $2\alpha_{\perp}$ for $U=0$. The critical SO coupling strength $2\alpha_c$, is shown in the inset of Fig.~\ref{fig:Uc}. The  behavior is similar to the critical Coulomb interaction $U_c$, however the slightly more linear functional dependence shows a modification of the spectrum due to SO coupling. 
At sufficiently large Coulomb interaction $U>U_c$ the width of the singlet-triplet transition as a function of $\epsilon$ in absence of SO (in other words the minimal distance between the two $\pi$-phase contributions) is further enhanced by a finite perpendicular component of the SO field to
$U+2\alpha_{\perp}$. 
 
 We point out that these effects inherent to the SO interaction depend on the angle between the SO interaction and the externally applied magnetic field. A rotation of the latter with respect  to the SO field would therefore allow to address these effects explicitly.  It is furthermore important to show that asymmetries between different dot states due to gating show features which are clearly distinguishable from the SO interaction.

\subsection{Asymmetry in coupling and detuning}\label{sec:asym}

\begin{figure}[t]
\centering
\includegraphics[width=3.5in]{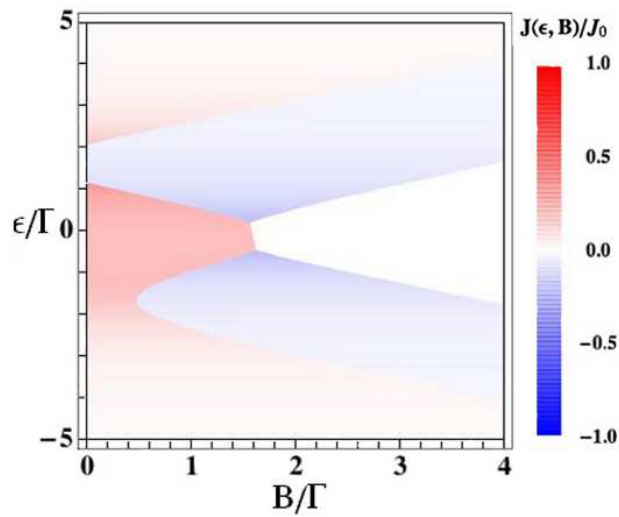}
\caption{(Color online) Josephson current for the  asymmetric case, $\beta=0.7$, $\delta=\Gamma$, as a function of the magnetic field and the mean level position in units of $\Gamma$ for $U=\Gamma$ ($U>U_c$) and $\phi=\pi/2$. The other parameters are $t=\Gamma$ and $k_B T=0$.
\label{fig:Jos_asym}}
\end{figure}

We consider, as a next step, the effects of asymmetries as generally present in experimental realizations of a double-dot setup on the features discussed before. The results are summarized in Figs.~\ref{fig:Uc} and \ref{fig:Jos_asym}. 

The asymmetry due to different couplings to the leads, $\beta\neq 0$, and due to a level detuning, $\delta\neq0$, result in changes in the $B$-field interval, given for the symmetric case in Eq.~(\ref{eq:B_sym_distance}), and consequently in  the critical Coulomb interaction, $U_c$. Since the expression for $\delta B$ in the asymmetric case is rather lengthy we do not present it here. However, we can extract the effect of   various asymmetries from the dependence of the critical interaction $U_c$ on the superconducting phase difference $\phi$, given in Fig.~\ref{fig:Uc}. A finite detuning $\delta \neq0$ still allows for a direct singlet-triplet transition at zero interaction if $\phi\rightarrow0$ and $\beta=0$. Importantly, for a pure coupling asymmetry, $\beta\neq0$ and $\delta=0$, a coexistence of the $0$-, the $\pi$- and the triplet-phase at zero Coulomb interaction $U=0$ is excluded and the required Coulomb interaction to allow for the singlet-triplet transition is strongly increased. This can be seen from the red dashed-dotted line in Fig.~\ref{fig:Uc}. In contrast, the figure shows that the overall effect of a finite detuning $\delta$ is to reduce the critical Coulomb interaction $U_c$.

We now concentrate on the situation, where the Coulomb interaction is larger than the critical value at which the singlet-triplet transition occurs, and study the effect of asymmetries on the Josephson current. 
An overall effect of the two types of asymmetries, $\delta\neq0$ and $\beta\neq0$, is a reduction of the absolute value of the  Josephson current.  A fundamental difference in the discontinuity at the
 singlet-triplet transition arises furthermore  when \textit{both} asymmetry effects are present at the same time, namely a  finite detuning and asymmetric dot-lead couplings, $\beta\neq0$ {\it and} $\delta\neq0$. 
In addition to a suppression of the Josephson current, it  exhibits a broken $\epsilon \to -\epsilon$ symmetry, as shown in Fig.~\ref{fig:Jos_asym}.
For the strong asymmetry chosen here, $\beta=0.7$ and $\delta=t$, a $\pi$-phase occurs at $B=0$ only for positive mean level energy $\epsilon>0$. We can understand this in the following way: For $\beta>0$, the left dot tends to be coupled more strongly than the right one; at the same time, for $\delta>0$, the single-particle level of the right dot is lowered in energy and hence its overlap with the bonding state is increased. When the gate voltage is chosen such that $\epsilon\approx \sqrt{\delta^2+t^2}/2$, half of the energy difference between the single-particle bonding and antibonding level, the bonding state (mainly given by the right dot state due to the detuning) tends to be occupied with a free spin and Josephson coupling through the antibonding state is enhanced (due to $\beta>0$). This leads to the $\pi$-phase contribution at $B=0$. However, when $\epsilon\approx - \sqrt{\delta^2+t^2}/2$ the antibonding state tends to have a free spin, the transport channel through the bonding state is more weakly coupled ($\beta>0$), and the $\pi$-phase only sets in at a finite $B$-field. The effect, shown in Fig.~\ref{fig:Jos_asym}, is reversed by changing either the sign of $\beta$ or of $\delta$.

This asymmetry goes along with yet a different effect, namely the tilting of the singlet-triplet transition line in correspondence of the discontinuity in $J$ as a function of $B$, resulting in an $\epsilon$-dependent $B_c$, differently to the symmetric situation.

In the case where both the effect of SO interaction and asymmetries occur, the two effects superimpose.

\subsection{Finite temperature effects}

\begin{figure}[b]
\centering
\includegraphics[width=3.in]{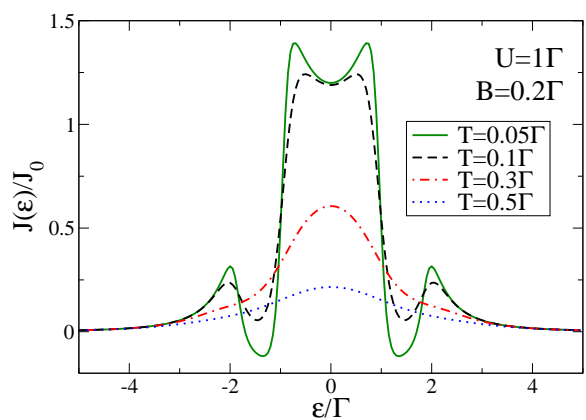}
\caption{(Color online) Josephson current at $B=0.2\Gamma$ as a function of $\epsilon$ in units of 
the coupling strength $\Gamma$ for different values of the temperature. The other parameters are $t=\Gamma$, $\phi=\pi/2$, $U=\Gamma$, $\beta=\delta=0$ and $\alpha=0$.
\label{fig:Joscut_epsilon_T}}
\end{figure}

\begin{figure}[t]
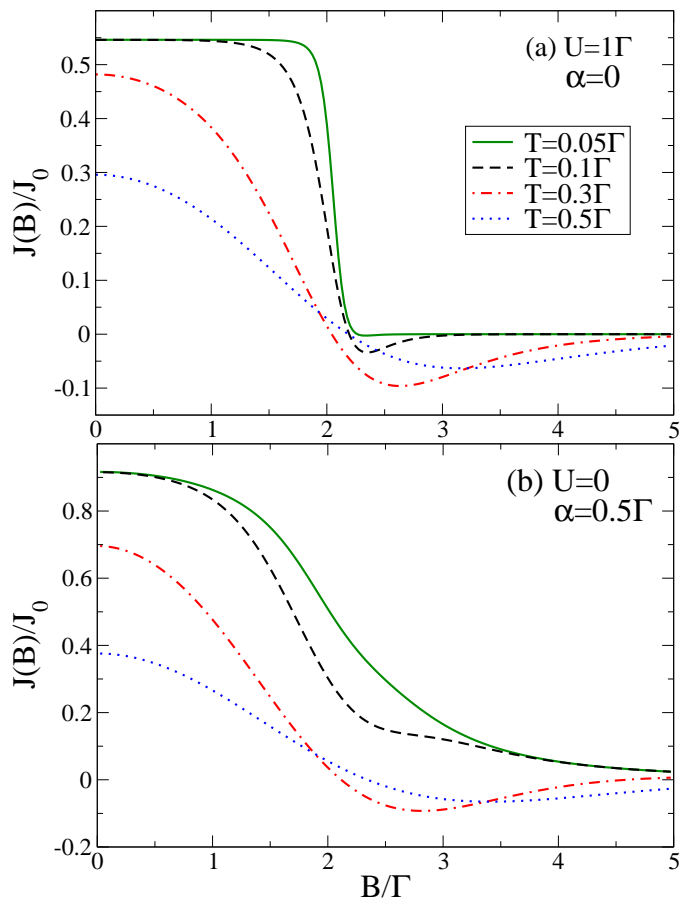

\centering
\includegraphics[width=3.5in]{fig9a.eps}
\includegraphics[width=3.5in]{fig9b}
\caption{(Color online) Josephson current for $\epsilon=0$  as a function of the magnetic field in units of 
the coupling strength $\Gamma$ for different values of the temperature. In (a) we 
set $\alpha=0$ and examine the case of finite Coulomb interaction $U=\Gamma$. In (b) we show the 
non-interacting case $U=0$ for finite SO interaction $\alpha=0.5\Gamma$ and $\theta=\pi/2$. 
The other parameters are $t=2\Gamma$, $\delta=\beta=0$, and $\phi=\pi/2$.  The reference current is $J_0(t=2\Gamma,\Gamma,\phi=\pi/2)$.
\label{fig:Joscut_T}}
\end{figure}

The analytic treatment of the effective Hamiltonian as outlined in  Sec.~\ref{sec:effective}, allows to compute the exact Josephson current, $J$, not only for $T=0$, but also for finite temperatures, see Eq.~(\ref{eq:Jos}). At finite temperatures, the transition between different phases is smeared out and the signal can be significantly reduced. Nevertheless, sign changes in the Josephson current have clearly been
detected experimentally~\cite{Cleuziou06,Maurand12,vanDam06}.
This reduction and smearing out of the signal due to finite temperatures is shown in Fig.~\ref{fig:Joscut_epsilon_T} for $\alpha=0$, where we show the Josephson current at $B=0.2\Gamma$ as a function of $\epsilon$ for different temperatures (see Fig.~\ref{fig:J_generic} for the corresponding $T=0$ results). We clearly observe that the effect of the temperature is a general reduction of the current. 
Most importantly, effects from the contribution of higher energy states are expected to modify the shape of the Josephson current significantly. 

In the following we focus on the Josephson current as a function of the magnetic field at $\epsilon=0$ to monitor the impact of finite temperatures $T>0$ on the singlet-triplet transition (at $\alpha_{\perp}=0$) and the smoothened crossover (at $\alpha_\perp\neq0$). In Fig.~\ref{fig:Joscut_T}, we show two representative situations  for the symmetric double-dot setup ($\beta=0$, $\delta=0$), namely the one for finite Coulomb interaction, $U>U_{c}$ without SO interaction ($\alpha=0$) in the upper panel~(a), and the one for finite orthogonal SO interaction, $\alpha_{\perp}=0.5 \Gamma$ and a vanishing Coulomb interaction $U=0$
 in the lower panel~(b). At $T=0$, for both parameter sets a direct singlet-triplet transition or respectively a crossover occurs, the
$\pi$-phases of the Josephson current setting in only at finite values of $\epsilon$. The reason for this is that the eigenstate of the system belonging to the subspace with one Nambu particle lies at higher energies for $\epsilon=0$, and can therefore not contribute to the Josephson current if there is not a large enough temperature allowing for the thermal occupation of higher-lying states.

We first discuss the effects at finite Coulomb interactions. In Fig.~\ref{fig:Joscut_T}~(a) the discontinuity in the Josephson current $J$ in correspondence of the singlet-triplet transition is gradually smeared out due to the increasing contribution of the higher energy states. Remarkably, in this regime the temperature \textit{stabilizes} the $\pi$-phase, while in general the temperature leads to a reduction of the $\pi$-phase, in terms of a narrowing of the regions of negative current as well as in lowering the absolute value. This reduction of the $\pi$-phase at finite temperature is visualized in the Josephson current as a function of the mean level energy $\epsilon$, at $B=0.2\Gamma$ far from the singlet-triplet transition, in Fig.~\ref{fig:Joscut_epsilon_T}, where indeed the temperature smears out the features in the current leading finally to a complete suppression of the $\pi$-phase. However, in contrast to the effect of the SO coupling leading to a smooth crossover instead of a sharp singlet-triplet transition, the temperature smears out \textit{all} transitions between different phases. 

A thermal stabilization effect of the $\pi$ phase  for a finite orthogonal SO interaction component with respect to the magnetic field $\alpha_{\perp}=0.5\Gamma$ is visible in Fig.~\ref{fig:Joscut_T}~(b), where the impact of the temperature on the Josephson current is reported for the non-interacting double quantum dot. The $\pi$-phase, which at $\epsilon=0$ and low temperatures is suppressed due to the SO interaction as in Fig.~\ref{fig:J_SOI}~(b), is recovered for sufficiently high temperatures. The required temperature for this effect to occur depends on the distance between the energy of the lowest lying states of the even and the odd Nambu particle sector respectively, which in turn is directly related to the value of $\alpha_{\bot}$, but also to the different tunnel-coupling strengths, $t$ and $\Gamma$. In the situation shown in Fig.~\ref{fig:Joscut_T} (b), the temperature recovering the $\pi$-phase is of the order of $\alpha_{\perp}$.
In this case, a \textit{sign change} of an initially \textit{completely positive} Josephson current $J>0$ at $\epsilon=0$ and $\phi=\pi/2$ is obtained due to a temperature increase only. For a system in the absence of SO interaction, $\alpha_\perp=0$, such an observation can only be expected in the presence of quasiparticle transport for a finite superconducting gap, where a small Josephson current  can flow in the triplet phase. 

\section{Parallel configuration}

\begin{figure}[b]
\centering
\center{\includegraphics[width=2.2in,clip=true]{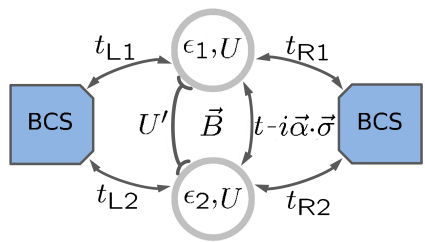}}
\caption{(Color online) Sketch of two parallel coupled quantum dots with parameters as in Fig.~\ref{fig:model}.
\label{fig:model1}
}
\end{figure}

Even though the physics of the serial double dot allows for a detailed insight in the physics of Josephson currents through quantum dots with SO interaction, also parallel quantum dots can be experimentally designed or can sometimes be a more appropriate model for a multilevel dot.
 We therefore consider in this section the situation of a double dot in the more general configuration containing also the parallel setup, as shown in Fig.~\ref{fig:model1}, with a finite 
coupling to both left and right superconducting lead for each dot,
\begin{eqnarray}
H_\mathrm{tun} = \sum_{i=\mathrm{L,R}}\sum_{j=1,2}\sum_{\sigma,k}t_{ij}d^{\dagger}_{j,\sigma} c^{}_{k\sigma i} +\mathrm{H.c.}\;~. 
\end{eqnarray}
We assume real and positive hopping amplitudes $t_{ij}$. The dot Hamiltonian and the lead Hamiltonian remain unchanged with respect to Eq.~(\ref{eq_Hdot}) and Eq.~(\ref{eq_Hleads}). 
In analogy to Sec.~\ref{sec:effective} (see also Appendix), we  derive an effective Hamiltonian, starting from the equations of motion for the double-dot Green's functions in the limit of an infinite superconducting gap.  Here, we find 
\begin{eqnarray}
&&H_\mathrm{eff}  =  H_\mathrm{dd}\\
&& -\left(\sqrt{\Gamma_{L1}\Gamma_{L2}}e^{i\phi/2}+\sqrt{\Gamma_{R1}\Gamma_{R2}}e^{-i\phi/2}\right)\left(d^{\dagger}_{1\uparrow}d^{\dagger}_{2\downarrow}+d^{\dagger}_{2\uparrow}d^{\dagger}_{1\downarrow}\right)\nonumber\\
&&-\left(\Gamma_{L1}e^{i\phi/2}+\Gamma_{R1}e^{-i\phi/2}\right)d^{\dagger}_{1\uparrow}d^{\dagger}_{1\downarrow}-\left(\Gamma_{L2}e^{i\phi/2}+\Gamma_{R2}e^{-i\phi/2}\right)d^{\dagger}_{2\uparrow}d^{\dagger}_{2\downarrow}+\mathrm{H.c.}\;~.\nonumber
\end{eqnarray}
 For $\Gamma_{L2}=\Gamma_{R1}=0$, the previous case of a serial double quantum dot is recovered. 
We observe that in contrast to the serial double  dot, a parallel configuration allows also for crossed Andreev reflection~\cite{Hofstetter09,Herrmann10,Melin03,Sauret05,Golubev07,Eldridge10}.
The physics of the parallel geometry is very rich due to interference effects. We here refrain from giving a systematic account of the various transport regimes (see Ref.~\cite{Meden06} for the classification of the different cases) and concentrate on one special situation which cannot be accessed by the serial geometry.

In the following we address one of the cases in which destructive interference leads to the complete decoupling of one of the molecular levels. This is the case for symmetric couplings, $\Gamma_{L1}=\Gamma_{L2}=\Gamma_{R1}=\Gamma_{R2}=\Gamma/2$, and vanishing detuning, $\delta=0$~\cite{Meden06,Boese01,Boese02,Karrasch06}, as can be seen 
performing a basis transformation of the effective (non-interacting) Hamiltonian to bonding and antibonding eigenstates of the isolated double dot
\begin{eqnarray}
\left|b,\sigma\right\rangle & = & \frac{1}{\sqrt{2}}\left(\left|1,\sigma\right\rangle+\left|2,\sigma\right\rangle\right) \nonumber \\
\left|a,\sigma\right\rangle & = & \frac{1}{\sqrt{2}}\left(\left|1,\sigma\right\rangle-\left|2,\sigma\right\rangle\right)\;.
\end{eqnarray}
It turns out that the antibonding state is completely decoupled from the superconducting 
leads and the tunnel Hamiltonian in this basis reads
\begin{eqnarray}
H_{\rm tun}=-\Gamma\left(e^{i\phi/2}+e^{-i\phi/2}\right)d^{\dagger}_{b\uparrow}d^{\dagger}_{b\downarrow}+{\rm H.c.}\;.
\end{eqnarray}
As a consequence, Cooper-pair transport is only possible through the bonding orbital.  

\begin{figure}[t]
\centering
\includegraphics[width=3.in]{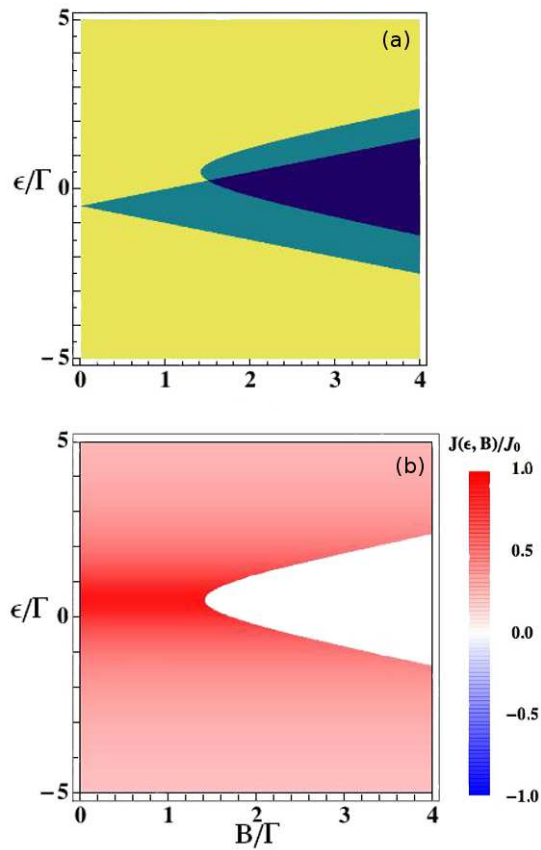}
\caption{(Color online) Phase diagram of the groundstate for a parallel geometry: singlet state (yellow), state with a free spin (green) and triplet state (dark blue) in (a), and Josephson current (b), for the symmetric, $\delta=\beta_{1}=\beta_{2}=0$, non-interacting case, $U=0$, as a 
function of the magnetic field and the level position in units of $\Gamma$. 
Furthermore $t=\Gamma$, $k_B T=0$, and $\phi=\pi/2$.
\label{fig:Parallel_U0}}
\end{figure}

We start with the discussion of the non-interacting dot system, $U=0$. 
Fig.~\ref{fig:Parallel_U0}~(a)  shows the phase diagram of the many-particle ground state
as a function of the gate voltage $\epsilon$ and the magnetic field $B$. The three different possible types of ground states are, as discussed in the previous section, a singlet state with Nambu particle number 2 (yellow), a triplet state with Nambu particle number 0  (dark blue) and a spin $1/2$ state with Nambu particle number 1 for spin $-1/2$ (green). 
The shape of the contributions of the spin $1/2$ states reflect that the system is composed of one effective level coupled to the superconducting leads (giving rise to the rounded green region) and a second effective level which is uncoupled (leading to the sharp triangular feature). This behavior is confirmed by the results for the Josephson current, see Fig.~\ref{fig:Parallel_U0}~(b).
For the Josephson current in this regime we find that indeed only the coupled bonding level contributes to the current. As a result we observe, that $J$ is either positive, $J>0$, or zero, $J=0$. The reason for this is obviously that the spin of the antibonding level is not coupled to the bonding level in this highly symmetric system. Therefore the Josephson current does not reveal a  $\pi$-phase as it was observed for the serial double dot  in the single-level transport situation. The transport through the parallel dot is blocked, whenever the coupled bonding level is occupied with a spin $1/2$ and hence Cooper pair transfer is not possible.

\begin{figure}[t]
\centering
\includegraphics[width=3.5in]{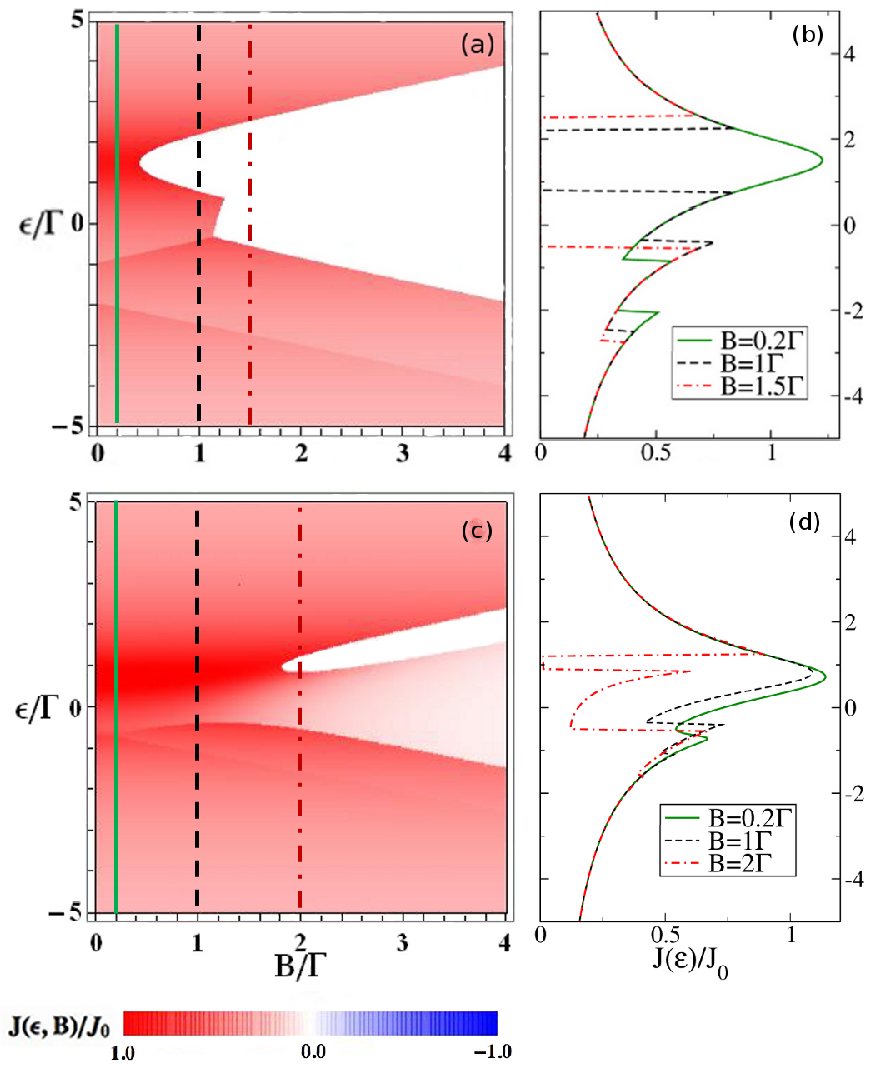}
\caption{(Color online) 
Density plot of the Josephson current for a parallel geometry, (a) for the interacting case, $U=\Gamma$ and $\alpha=0$, and (c) for a finite SO coupling $\alpha_\perp=0.5\Gamma$ in the noninteracting case, $U=0$. The current is shown as a 
function of the magnetic field and the level position in units of $\Gamma$. (b) and (d) cuts through the density plots in (a) and (c) of the Josephson current as a function of $\epsilon$ and fixed values of $B$. The other parameters are the same as in Fig.~\ref{fig:Parallel_U0}.
\label{fig:Parallel_U1}}
\end{figure}

As in the serial setup we use the situation of a finite Coulomb interaction as the reference situation to study SO interaction effects. The effect of Coulomb interaction is summarized in  Fig.~\ref{fig:Parallel_U1}~(a-b). 
As for the non-interacting case, the decoupling of the antibonding state is reflected in a broken $\epsilon\rightarrow -\epsilon$ symmetry. In presence of a finite Coulomb interaction, this broken symmetry in addition appears in a tilted singlet-triplet transition line of the ground state.
The Josephson current shows that only a single level supports a Josephson current, while transport through the other level is blocked.  Also the magnetic moment of the uncoupled level, when singly occupied, does not lead to a $\pi$-phase in the Josephson current through the other level.  
In contrast to the noninteracting case, the Josephson current does however change its amplitude when the ground state changes between a singlet and the doublet state, which can here be observed at negative $\epsilon$. The latter is due to a capacitive coupling between the two levels entering through the interaction term $U$.

\subsection{Spin-orbit interaction effects}

Also in the parallel setup the effects of the SO interaction can be clearly pointed out. We  discuss the influence of the SO interaction on the Josephson current through the parallel double-dot setup, when the effective SO field is perpendicular to the externally applied magnetic field. The results are shown in Fig.~\ref{fig:Parallel_U1} (c-d). For simplicity we consider $U=0$, the physical behavior is qualitatively not affected by a finite Coulomb interaction. Three different effects are visible due to the SO coupling: (i) it has a similar effect on the current as the Coulomb interaction in the sense that no state of spin $\pm 1/2$ is accessible in the vicinity of $\epsilon\approx0$, allowing for a direct transition between a singlet and a triplet state. (ii) In contrast to the case without SO coupling, the discontinuous transition between positive and suppressed current is replaced by a smooth crossover as it was observed in the serial double-dot setup. This smooth transition again originates from the coupling between the singlet and the triplet many-body ground state. This effect makes the regions in which the Josephson current tends to be blocked distinguishable. The reason is that in the subspace with one Nambu particle no current can flow through the bonding level when it is singly occupied. However, when the transporting bonding level tends to be blocked with one spin and at the same time also the antibonding level is, then SO coupling induces coupling to the current carrying singlet state. Therefore the region in which the current is fully blocked becomes much smaller (as expected for $U=0$ when comparing to the small $\pi$-phase in the noninteracting serial double dot studied before). (iii)  We observe the absence of a $\pi$-phase originating from the fact that only a single level is effectively coupled to the superconducting leads being insensitive to the magnetic moment of the uncoupled level. The Josephson current however appears to be sensitive to the occupation of the (uncoupled) antibonding level \textit{through the SO coupling}, which leads to discontinuities in the current, between the regions where the antibonding level tends to be occupied or empty.

\section{Conclusions}
\label{concl}

We studied the effect of SO interaction $\alpha$ on the equilibrium supercurrent through a serially coupled double quantum dot as a minimal model to describe a multi-orbital system, attached to superconducting leads. The SO effects  were illustrated in a study of the Josephson current and contrasted with effects due to Coulomb interaction and  of coupling asymmetries.

We find a pronounced dependence on the orientation of an applied magnetic field with respect to the spin-orbit interaction: for $\boldsymbol{\alpha} || \boldsymbol{B}$, the SO interaction does not qualitatively affect the singlet-triplet transition.
However, for a finite orthogonal component, $\alpha_\perp$, this transition is smoothened as the singlet-triplet transition is inhibited by an anticrossing of the corresponding levels. 
Extending the study to finite temperatures we find that $\alpha_\perp$ has a similar effect as the Coulomb interaction in the stabilization of a $\pi$-phase arising around zero gate voltage.
We finally discuss the Josephson current for the situation of an effective decoupling of double-dot states, which occurs for a symmetric parallel dot configuration, and point out the effect of SO interaction in this more general setup. 
Our detailed analysis can serve as a basis for an extended study including quasi-particle transport due to a finite superconducting gap or through hybrid structures involving superconducting as well as normal leads, thereby allowing for an inclusion of non-equilibrium effects.  

\ack We thank Christoph Karrasch,  Volker Meden,  Federica Haupt and Michele Governale  for helpful discussions. J.~S. and S.~D. acknowledge financial support from the Ministry of Innovation NRW, and S.~A. from the Deutsche Forschungsgemeinschaft (FOR 923).

\appendix
\section{Derivation of the effective Hamiltonian of the serial double quantum dot}

The effective Hamiltonian can be derived by setting up an equation of motion for the full double-dot Green's function and showing the equivalence with the equation of motion of an effective Hamiltonian. The  retarded Green's function, which we use for this purpose, is defined as $\hat{G}^\mathrm{ret}_{ji}(t)=-i\theta(t)\langle \left\{\Psi_{i}(t),\Psi_{j}(0) \right\}\rangle$, where the Nambu spinor $\Psi_{j}=(d^{}_{j\uparrow },d^{\dagger}_{j\downarrow})$ has been introduced. The $\theta$-function assures that retarded processes are accounted for. Due to the vector form of the spinors, the Green's function is a $2\times 2$ matrix for electron and hole contributions. Off-diagonal elements appear due to the coupling of electrons and holes by the superconducting leads. 
We start by expressing the microscopic Hamiltonian of the full superconductor - double-dot - superconductor system, introduced in Sec.~\ref{sec:model} in terms of Nambu spinors. At first we can omit the Coulomb interaction and the SO part for sake of clarity and show in the following that only terms in the Hamiltonian involving lead operators are affected by the coupling to the superconductor on the level of the Hamiltonian. Therefore the terms $H_\mathrm{int}$ and $H_\mathrm{SO}$ are independent contributions which can simply be added to the effective Hamiltonian in the end of the procedure. Note however, that the results have to be modified if both dots couple to both leads as it is the case in the parallel setup. 
We start our evaluation from the microscopic Hamiltonian
\begin{eqnarray}\label{eq:H_Nambu}
\tilde{H} & = & H_{0}+H_\mathrm{lead}+H_{T}\\
& = & \sum_{ j=1,2}\left[\Psi^{\dagger}_{j}\hat{H}_{{ j}}\Psi^{}_{j}+\Psi^{\dagger}_{j}\hat{H}_{t}\Psi^{}_{\overline{j}}\right]+\sum_{k,s}\Psi^{\dagger}_{ks}\hat{H}_{k}\Psi^{}_{ks}\nonumber \\
 &&  +\sum_{k}\left(\Psi^{\dagger}_{1}\hat{H}_{T_{L}}\Psi^{}_{kL}+\Psi^{\dagger}_{2}\hat{H}_{T_{R}}\Psi^{}_{kR}+{\rm H.c.}\right)\nonumber
\end{eqnarray}
where $\Psi_{ks}=(c_{sk\uparrow },c^{\dagger}_{s-k\downarrow })$ is the spinor describing the lead operators and $\bar{j}$ equals 1 if $j=2$ and vice versa. The Hamiltonian contributions in matrix form are explicitly given by
\begin{eqnarray}\label{eq:H_matrices}
\hat{H}_{ j} =
\left(\begin{array}{cc}
\epsilon_{j,\uparrow} & 0 \\
0 & -\epsilon_{j,\downarrow} 
\end{array}\right) & , & 
\hat{H}_{t}=
\left(\begin{array}{cc}
-\frac{t}{2} & 0 \\
0 & \frac{t}{2} 
\end{array}\right)\\
\hat{H}_{k} =
\left(\begin{array}{cc}
\epsilon_{k} & -\Delta \\
-\Delta & -\epsilon_{k} 
\end{array}\right) & ,  &
\hat{H}_{T_{s}}=
\left(\begin{array}{cc}
t_{s} e^{i \frac{\phi_{s}}{2}} & 0 \\
0 &  -t_{s}e^{-i\frac{\phi_{s}}{2}}
\end{array}\right).
\nonumber
\end{eqnarray}
We here performed a transformation, which absorbs the phase of the superconducting gap into the tunnel matrix elements, $c_{sk\sigma}\rightarrow c_{sk\sigma}e^{i\phi_s/2}$. 

The equation of motion is set up by considering the matrix equation $i\frac{d}{dt}\hat{G}^\mathrm{ret}_{ji}(t)=\delta(t)-i\theta(t)\langle\left\{-\left[\tilde{H},\Psi_i(t)\right],\Psi_j(0)\right\}\rangle$, see e.g. Ref.~\cite{Bruus}. 
The resulting set of equations simplifies after a Fourier transform, with $G_{ji}:=G_{ji}(\omega)=\int dt e^{-i\omega t}G_{ji}(t)$, yielding
\begin{eqnarray}
\left(
\begin{array}{cc}
\omega -\hat{H}_{1}-\hat{H}_\mathrm{L} & -\hat{H}_{t}\\
 -\hat{H}_{t} &  \omega -\hat{H}_{2}-\hat{H}_\mathrm{R}
\end{array}
\right)
\left(
\begin{array}{cc}
G_{11} & G_{12}\\
G_{21} & G_{22}
\end{array}
\right) & = & 1\ .\nonumber\\ \label{eq:Greenfct}
\end{eqnarray}
The expressions $H_s$ with $s=\mathrm{L,R}$  introduced in Eq.~(\ref{eq:Greenfct}) 
include the effect of the superconducting leads 
\begin{equation}
\hat{H}_{s}:=\sum_{k}\hat{H}_{T_{s}}
\left(\omega-\hat{H}_{k}\right)^{-1}\hat{H}^{\dagger}_{T_{s}}  \ .
\label{eq:H_tun}
\end{equation}
They show that the $\Delta$-dependence affects only those parts of the  Hamiltonian  which describe the coupling between the leads and the dots. Therefore we have to take the limit $\Delta\rightarrow\infty$ only in $\hat{H}_{s}$. We here consider the wide-band limit, where the density of states in the leads in their normal conducting state is constant. We can then write the sum over the wave vector $k$ as an integral and obtain
\begin{eqnarray}
\hat{H}_{s} & = & \int^{\infty}_{-\infty}d\epsilon\frac{\rho_{0}\left|t_s\right|^2}{\omega^{2}-\epsilon^{2}-\Delta^{2}}\left(\begin{array}{cc} \omega+\epsilon& \Delta e^{i\phi_{s}} \\ \Delta e^{-i\phi_{s}}& \omega-\epsilon \end{array}\right) \nonumber 
\end{eqnarray}
\begin{eqnarray}
&& \quad= \frac{\Gamma_s}{\sqrt{\Delta^{2}-\omega^{2}}}\left(\begin{array}{cc} -\omega & -\Delta e^{i\phi_{s}} \\ -\Delta e^{-i\phi_{s}} & -\omega \end{array}\right)\ .
\label{eq:sum_k}
\end{eqnarray}
Performing the limit  $\Delta\rightarrow\infty$ results in
\begin{equation}
\hat{H}_{s}^\mathrm{eff}=\left(\begin{array}{cc}
0& -\Gamma_{s}e^{i\phi_{s}}  \\
 -\Gamma_{s}e^{-i\phi_{s}} & 0 
\end{array}\right)\ .
\label{eq:H_i_final}
\end{equation}
The effective Hamiltonian in the limit of  $\Delta\rightarrow\infty$ replaces the one given in  Eq. \eref{eq:H_Nambu} and is given by
\begin{equation}\label{eq:H_Nambu_eff}
\tilde{H}_\mathrm{eff} = \sum_{ j}\left[\Psi^{\dagger}_{j}\hat{H}_{j}\Psi^{}_{j}+\Psi^{\dagger}_{j}\hat{H}_{t}\Psi^{}_{\overline{j}}+\sum_{s}\Psi^{\dagger}_{j}\hat{H}^\mathrm{eff}_{s}\Psi^{}_{j}\right]\/ .\\
\end{equation}
Transforming Eq. \eref{eq:H_Nambu_eff} back into the electron basis and adding the Coulomb and SO interaction  results in the effective Hamiltonian for the serial double quantum dot given in Eq. \eref{eq:hatom}. 
\\
\\

\providecommand{\newblock}{}

\end{document}